# Energy and material efficient non-circular bore Bitter magnets

A. Akhmeteli[1], A.V. Gavrilin[2]

*Abstract* – **There exist a number of experiments/applications where the second dimension of the bore of Bitter magnets is not fully utilized. Using an analytical solution for elliptical bore coils, we show that reducing one of the dimensions of the bore can lead to considerable decrease in consumed power and/or coil material. The results of additional numerical computation show that a 20% higher magnetic field can be achieved for an elliptical bore and racetrack bore coils with 1:10 ratio of the transverse dimensions of the bore than for a circular bore coil with the same larger transverse dimension of the bore, consumed power, and disk area.**

## 1. Introduction

It is common knowledge that Bitter magnets, made from copper alloy sheet metal (so-called Bitter disks), produce continuous magnetic fields significantly above 30T. The up-to-date Florida-Bitter technology developed at the National High Magnetic Field Laboratory, Tallahassee, Florida, enables one to build 40-T class magnets using about 28MW of power [1], which is a real progress from the standpoint of energy consumption reduction (in terms of MW per Tesla) at the top range of field available for experimentalists/users presently.

At the same time, even 28 MW of power is still a lot, and so an active search for avenues for reduction of power consumed by Bitter magnets continues. In this article, we would like to suggest a different line of attack on the problem of energy bill reduction. Also, a considerable gain in Bitter disk material, which is expensive and expendable, can be obtained by using an approach discussed below.

Traditionally, a Bitter magnet bore is of perfectly round shape, a circle, i.e., a geometrical figure having 2 dimensions. Do magnet users always utilize both of these dimensions? Do they need badly the entire bore? The answer seems to be the following: there exist a number of experiments/applications where the second dimension is not really needed and not actually used, whereas some other experiments can be so "flatly" designed that the second dimension is not utilized or is utilized very little. In other words, the idea that we would like to discuss here lies in significant reduction of the second dimension to save on power and/or material. One of the practical solutions may turn out to be a Bitter magnet with an elliptic bore and correspondingly elliptic outer surface (Fig. 1). An elliptic bore magnet is the first choice to examine, because there is an analytical solution for a Bitter disk whose outer and inner boundaries are homothetic ellipses [2-4]; there also exists an analytical solution for a Bitter disk with confocal elliptical outer and inner boundaries as well (see [4]). In [2-4], these solutions were used to calculate the magnetic field generated by a so-called tilted (canted) coil assembled from elliptic Bitter disks at an angle to the coil axis to form the circular bore [5,6]; the very first conceptual design of a Bitter magnet consisted of such coils was suggested by M.D. Bird [6]. However, the same mathematical apparatus can be also employed to calculate the magnetic field of an elliptic bore Bitter coil assembled from elliptic disks at a 90-degree angle w.r.t. the coil axis.

---

[1] LTASolid Inc., 10616 Meadowglen Ln 2708, Houston, TX, USA, akhmeteli@ltasolid.com
[2] National High Magnetic Field Laboratory – FSU, Tallahassee, FL 32310, USA; gavrilin@magnet.fsu.edu

## 2. Magnetic field. Analytical solution

We have made some estimates for magnets with more or less typical dimensions.

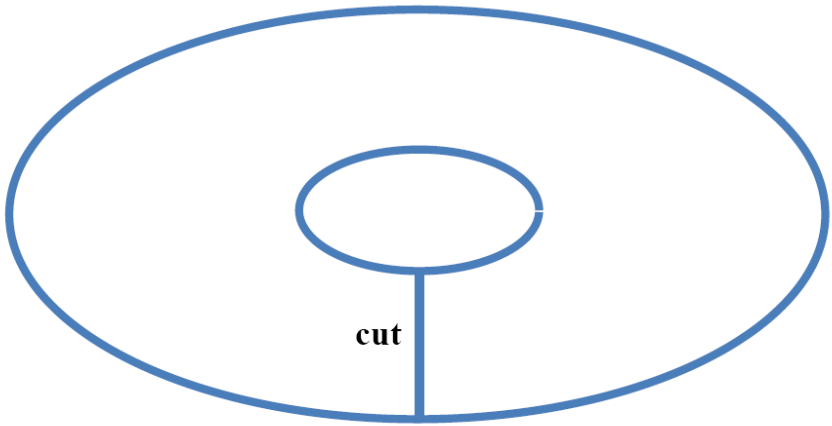

Fig. 1. Schematic of an elliptic Bitter disk / cross-section of an elliptic bore Bitter coil. The inner and outer ellipses are homothetic, with the short semi-axes significantly smaller than the long semi-axes. The cooling holes are not shown.

Example 1. Let us consider a single coil Bitter magnet with a circular bore 35 mm in inner radius and 100 mm in outer radius and compare it with an elliptic bore single coil Bitter magnet being equal in power consumed to generate the same on-axis magnetic field, albeit using less material. The inner ellipse (the bore) semi-major and semi-minor axes of the "elliptic" magnet are 35 mm and 17.5 mm, respectively, whereas the outer ellipse semi-major and semi-minor axes were chosen to be equal to 119.7 mm and 59.85 mm, respectively, to obtain the same on-axis field and to make the disk inner and outer ellipses homothetic (Fig. 2). Thus, the second dimension of the bore is reduced by 50% (while the first one is intact). This results in the material reduction by 25%, although the elliptic bore coil turns out to be slightly thicker along the disk's major axis than the circular bore coil.

Example 2. Let us compare the same (as above) circular bore Bitter coil with an elliptic bore Bitter coil being equal to the former in the amount of material (of Bitter disks) used at the same on-axis field, hoping that such a coil will use less power. The elliptic bore of this coil is the same as the elliptic bore coil has in Example 1 (the bore second dimension is halved), and the outer ellipse was chosen to have the semi-major and semi-minor axes 137.022 mm and 68.61 mm, respectively (so as to keep the material amount constant). Indeed, it turns out that this elliptic bore coil needs 11% less power than the circular bore coil does.

As can be inferred from the examples, even a rather moderate reduction in the bore's second dimension can result in noticeable gain in either power or material used, or both. This gain may turn out to be more impressive if one further reduces the second dimension. Presumably, the saving on power may exceed 25% at greater (than 2), albeit still technologically and practically reasonable, ratios between the semi-major and semi-minor axes. Obtainment of more precise values invites a comprehensive analysis that can be done later.

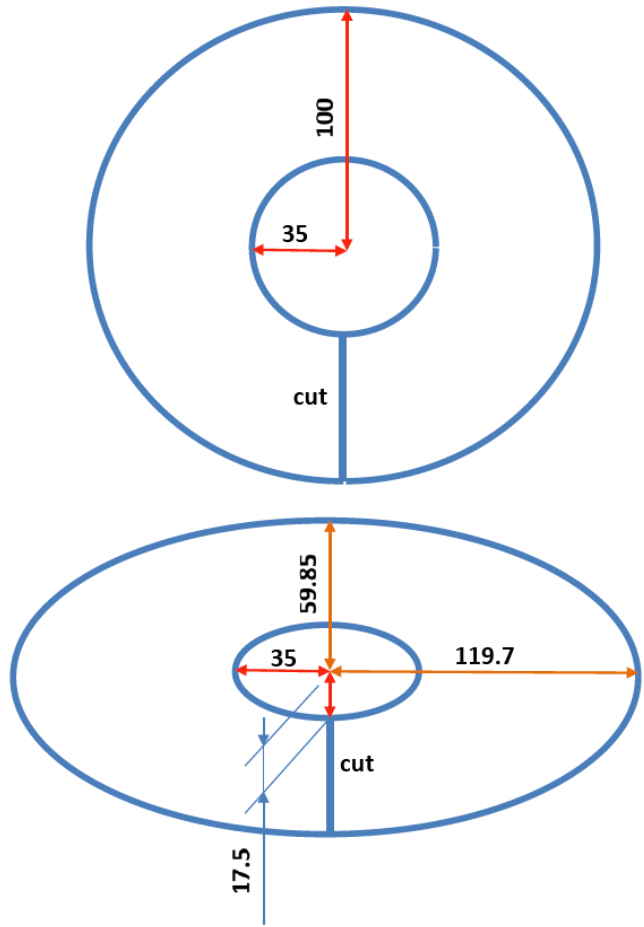

Fig. 2. Schematic and dimensions of Bitter disks of circular bore (upper) and elliptic bore (lower) Bitter coils from Example 1.

There exist very important structural mechanical aspects that should be studied to complete the picture. Let us start with the following. As can be inferred from Fig. 3, the body forces in the elliptic coil (from Example 1) are distributed in a different and more complicated way compared to that in the round coil. The force is the highest in value in the area where the radius of curvature is the smallest (on the major axis), and this force is noticeably higher than that in the round coil.

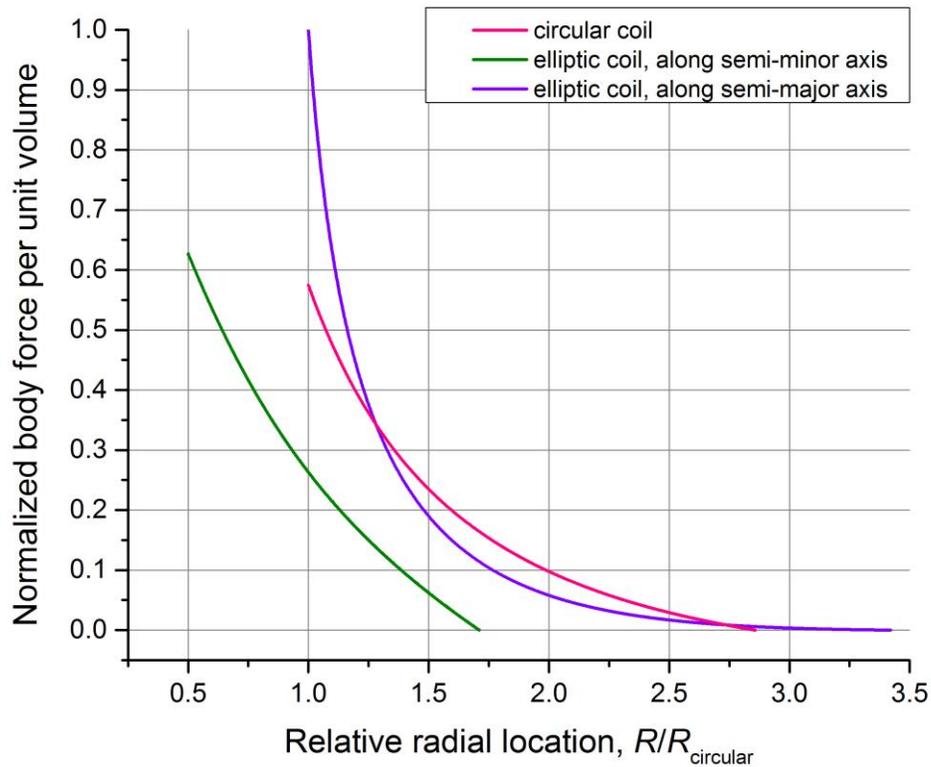

Fig. 3. Comparison of body force per unit volume in the elliptic (Example 1) and round coils.

## 3. Stress. Numerical computation

We performed computations of stress in the elliptic bore coil and circular bore coil of Example 1 using plane stress finite element analysis. The boundary of the elliptic bore coil without current (no magnetic field) versus the deformed mesh of the same coil with a current is shown in Figure 4. The displacement is in arbitrary units, the Poisson ratio is 0.33.

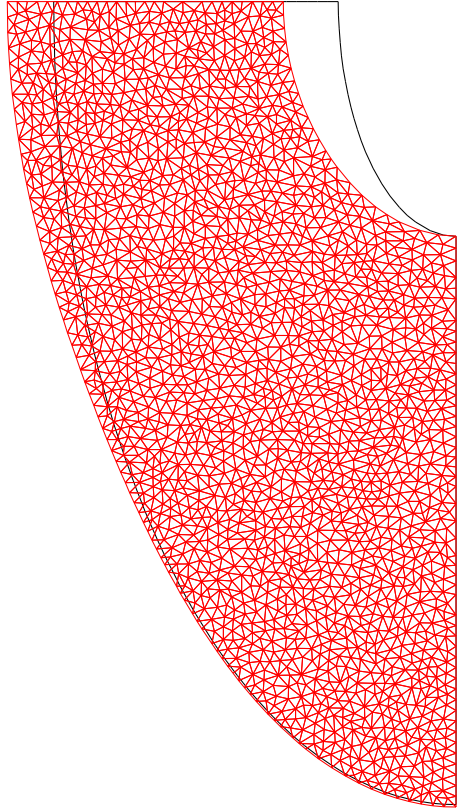

Fig. 4. The boundary of one fourth of an elliptic bore coil with no current versus the deformed mesh under current, with no support. Rotated.

The contour plot of von Mises stress is shown in Fig. 5.

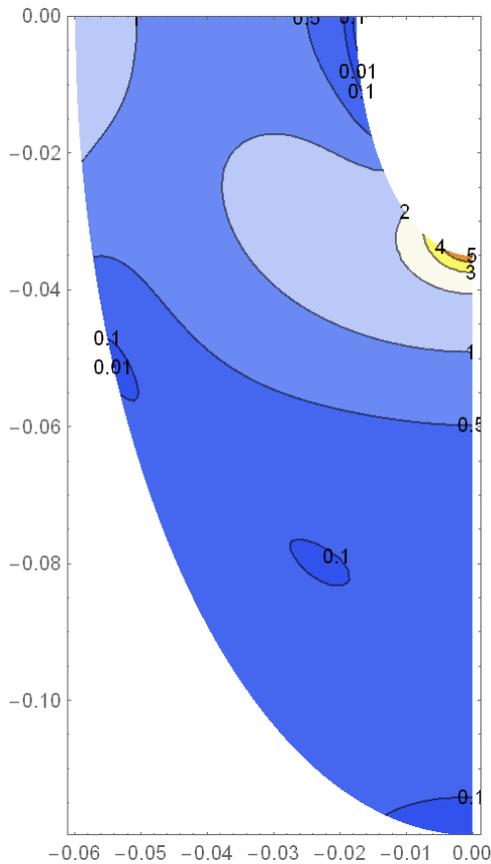

Figure 5. Contour plot of von Mises stress, one fourth of an elliptic bore coil, with no support. Rotated. The inscriptions are the stress in arbitrary units.

The maximum von Mises stress was 3.18 times higher for the elliptic bore coil than that for the circular bore coil, evidently because of the presence of bending moments. In order to address the issue, we mimicked a support (overlapping) counteracting the moments (in the elliptic bore coil) as follows: the displacements were assumed to vanish on the part of the external elliptic boundary that is symmetric with respect to the minor axis of the ellipse and has a length of projection on the major axis equal to the semi-major axis (of the external ellipse). The boundary of the elliptic bore coil without current (no magnetic field) versus the deformed mesh of the same coil with a current is shown in Figure 6.

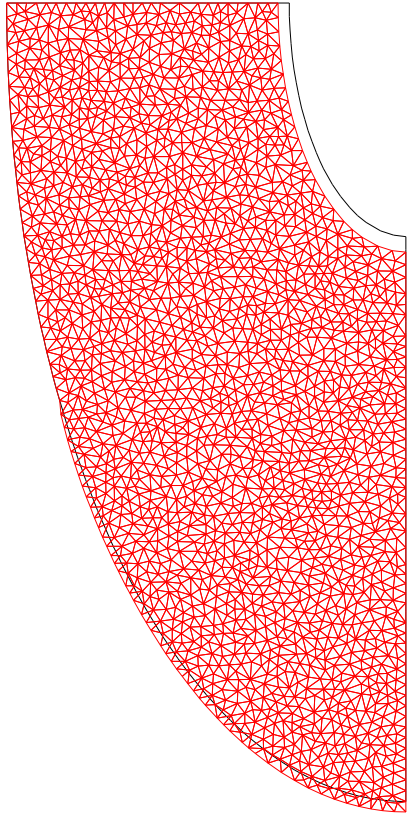

Fig. 6. The boundary of one fourth of an elliptic bore coil with no current versus the deformed mesh under current, with support. Rotated.

The contour plot of von Mises stress is shown in Fig. 7.

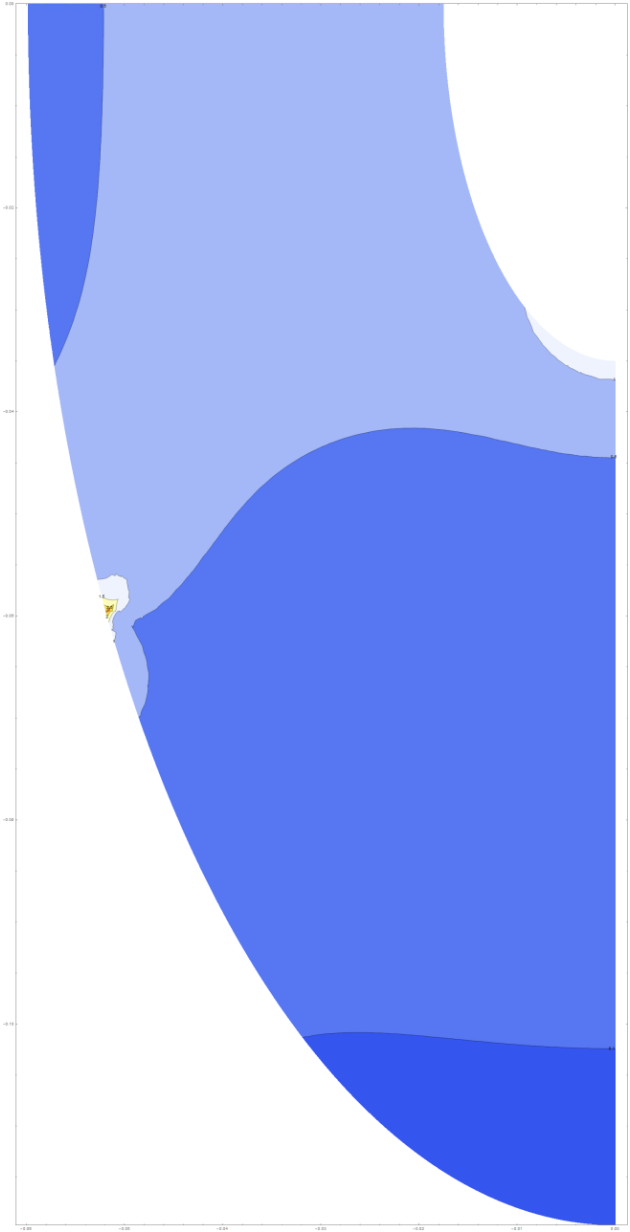

Figure 7. Contour plot of von Mises stress, one fourth of an elliptic bore coil, with support. Rotated.

The maximum von Mises stress turned out to be only 1.67 times higher than that in the circular bore coil (without a support) and was concentrated at the ends of the support. Elsewhere the von Mises stress was smaller than the maximum von Mises stress in the circular bore coil (the maximum factor was 0.85), and this is likely the result, a quite encouraging one, that we need. The high stress at the ends of the support is largely an artifact of the crude model of the support: on the one hand, the support was assumed to be completely "glued" to the coil; on the other hand, the ends of the support were not rounded. The stress would be noticeably lower if contact elements were used and the support size and shape were optimized, as that would lead to better stress distribution and thus to reduction of the maximum stress. In addition, we realize that a real high field Bitter magnet consists of multiple nested coils with rather limited gaps between them, which complicates placement of external support structures to some extent, so the tie rods properly distributed over the coil volume nearby the minor axis, where the current density is relatively low, could counteract the bending moments in some or significant degree to help the external supports, if any. Anyway, although the obtained results are promising, they are preliminary and invite further analysis, clarification, and refinement. It is also worthy of notice that optimization of cooling holes in the elliptic bore coils may turn out to be somewhat more complicated and require a larger number of iterations compared to the circular bore coils. Also, in an elliptic bore coil, each disk cut location is supposed to be calculated carefully, since the disks are not of round shape.

We do not claim that the elliptic bore is the best choice, because a somewhat different shape may turn out to be more optimal and thus beneficial or preferable in terms of strength of the magnet (there may be some trade-offs). Particularly, the stress can be probably managed better if the ellipses are replaced with racetrack curves. Indeed, in the racetrack curves, the curvature is uniform (where it does not vanish), which may result in lower stress. Nonetheless, the concept practicability can depend on a number of factors.

## 4. Magnetic field. Numerical computation

The currently available implementation of the analytical solution has limitations. For example, it needs some modifications for calculation of magnetic field for an elliptic bore with high eccentricity, and such calculation is desirable to get an idea of the potential of the optimization proposed in this work. Furthermore, the analytical solution is not applicable to non-homothetic inner and outer ellipses or other shapes of the disk. Therefore, we used a numerical computation of the magnetic field. The Laplace problem for potential was solved for a quarter disk using the finite element method. Using the symmetry of the problem, the Dirichlet conditions with constant potentials were posed at the straight boundaries of the quarter disk and the Neumann conditions at the curvilinear boundaries. The current density equals the gradient of the potential up to a factor and was computed using numerical differentiation; and the component of the magnetic field along the axis of the (infinite) magnet was computed by integration of the current density using the curl theorem. The accuracy of the numerical computation method was checked by comparison with the results of the analytical solution for Example 2 above.

In the following examples, the ratio of the transverse dimensions of the bore is 1:10.

Example 3. We compared a single coil Bitter magnet with a circular bore 35 mm in inner radius and 100 mm in outer radius with an elliptic bore single coil Bitter magnet with the same power consumed, disk area and larger dimension of the bore. The inner ellipse (the bore) semi-major and semi-minor axes of the "elliptic" magnet are 35 mm and 3.5 mm, respectively, whereas the outer ellipse semi-major and semi-minor axes were chosen to be equal to 100 mm and 89 mm, so the inner and outer ellipses are not homothetic (Fig. 8). The magnetic field is 23% higher for the elliptic bore coil magnet. When we tried to change the major and minor semi-axes of the outer ellipse without changing the area of the disk, we did not obtain better results.

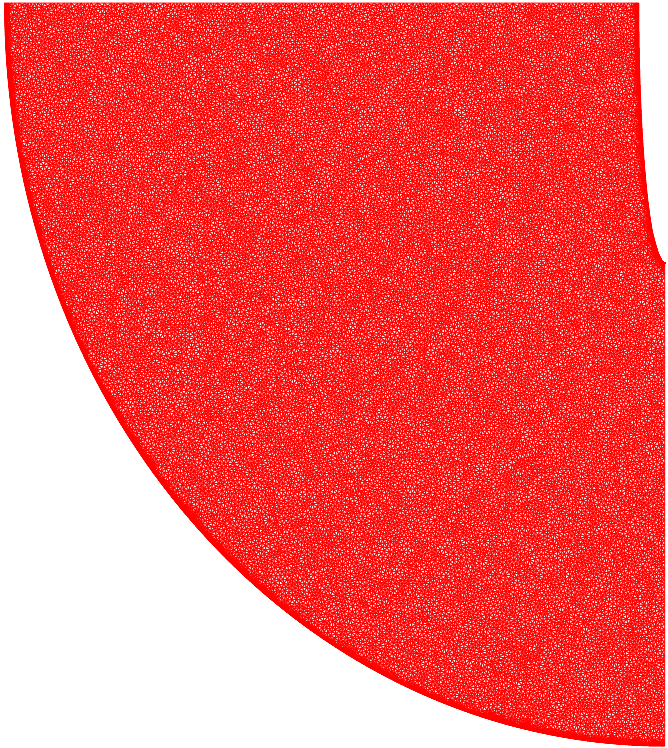

Figure 8. The fine mesh for the elliptic bore coil magnet of Example 3.

Example 4. We compared a single coil Bitter magnet with a circular bore 35 mm in inner radius and 100 mm in outer radius with a racetrack bore single coil Bitter magnet with the same power consumed, disk area and larger dimension of the bore. The radii of the turns of the outer and inner curves of the racetrack are 76.5 mm and 3.5 mm, respectively, whereas the length of the straight segments of the racetrack is 63 mm (Fig. 9).
The magnetic field is 20% higher for the racetrack bore coil magnet. This is slightly less than for the elliptic bore coil magnet of Example 3, simply because the racetrack bore cross-section area is somewhat larger than the elliptic bore one. In other words, the difference obtained does not necessarily mean that the elliptic bore coil is slightly more efficient. Also, the shape of the racetrack bore may be preferable for some applications, and the racetrack-shaped disks may turn out to be easier to produce and stack.

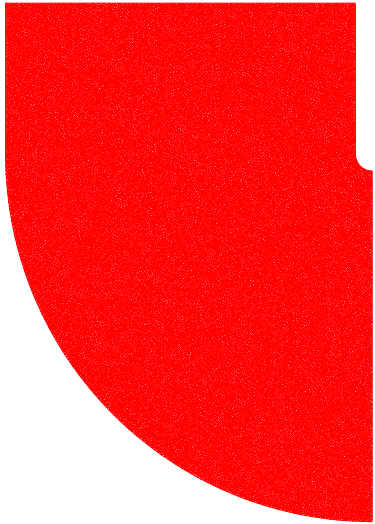

Figure 9. The fine mesh for the racetrack bore coil magnet of Example 4.

We have not performed stress computation for the magnets of Examples 3 and 4, but we expect the results to be similar to those of Section 3 and quite promising.

In closing, the possibility should not be ruled out that the concept can be extended to wire-wound superconducting magnets with a view to saving on the conductor and/or reducing the stress due to Lorentz forces. However, right now we do not know if it makes sense or not: further research could clarify the situation. Also, we are aware of the fact that wire-wound non-round coils were considered and built, however, nobody did it yet, aiming to save on the materials or to increase magnetic field at the same stored energy: the shape choice was dictated by the applications mostly, if not only.

**Acknowledgments**

The authors are grateful to Mark D. Bird for interest in this work, important critical remarks and timely advice, and to Jack Toth for fruitful discussions of complex optimization in Bitter magnets.

**References**

1. https://nationalmaglab.org/magnet-development/magnet-science-technology/magnet-projects/28-mw-magnet
2. A.M. Akhmeteli and A.V. Gavrilin, "Superconducting and Resistive Tilted Coil Magnets," arXiv:physics/0410002 (2004)

3. A.M. Akhmeteli, A.V. Gavrilin, and W.S. Marshall, "Superconducting and Resistive Tilted Coil Magnets for Generation of High and Uniform Transverse Magnetic Field", *IEEE Trans. Appl. Supercond.*, 15, 2, 1439-1443 (2005)
4. A.M. Akhmeteli, A.V. Gavrilin, and W.S. Marshall, "Superconducting and Resistive Tilted Coil Magnets. Magnetic and Mechanical Aspects", *chapter 5 in book "Superconductivity, Magnetism and Magnets"*, ed. L.K. Tran, Nova Publishers, ISBN: 1-59454-845-5, 139-172 (2006)
5. A.V. Gavrilin and M.D. Bird, "Transverse Field Bitter-Type Magnet" U.S. Patent No. 6,876,288, Awarded April 05 (2005)
6. A.V. Gavrilin, M.D. Bird, S.T. Bole, and Y.M. Eyssa, "Conceptual Design of High Transverse Field Magnets at the NHMFL," *IEEE Trans. Appl. Supercond.*, 12, 1, 465-469 (2002).